\documentclass{llncs}

\usepackage{amssymb}
\usepackage{color}
\usepackage{amsmath}
\usepackage{graphicx}
\usepackage{caption}
\usepackage{subcaption}
\usepackage{wrapfig}
\PassOptionsToPackage{hyphens}{url}\usepackage{hyperref}
\usepackage{enumitem}

\definecolor{orange}{rgb}{1,0.5,0}

\newcommand{\done}[1]{}

\newcommand{\doublecell}[2][c]{%
  \begin{tabular}[#1]{@{}l@{}}#2\end{tabular}}

\urldef{\mailsa}\path|{wuechner, ochoa, pretschn}@cs.tum.edu|

\begin{document}

\title{Robust and Effective Malware Detection through Quantitative Data Flow Graph Metrics}

\author{Tobias W\"uchner\and Mart\'in Ochoa\and Alexander Pretschner}
\institute{Technische Universit\"at M\"unchen, Germany
}

\maketitle

\begin{abstract}
We present a novel malware detection approach based on metrics over quantitative data flow graphs. 
Quantitative data flow graphs (QDFGs) model process behavior by interpreting issued system calls as aggregations of quantifiable data flows.
Due to the high abstraction level we consider QDFG metric based detection more robust against typical behavior obfuscation like bogus call injection or call reordering than other common behavioral models that base on raw system calls. We support this claim with experiments on obfuscated malware logs and demonstrate the superior obfuscation robustness in comparison to detection using n-grams.
Our evaluations on a large and diverse data set consisting of about 7000 malware and 500 goodware samples show an average detection rate of 98.01\% and a false positive rate of 0.48\%. Moreover, we show that our approach is able to detect new malware (i.e. samples from malware families not included in the training set) and that the consideration of quantities in itself significantly improves detection precision.
\end{abstract}

\section{Introduction}
\label{sec:introduction}

Despite the increasing availability and deployment of intrusion detection 
systems and anti-virus engines, malicious software (malware) remains a severe 
threat. One reason is the steadily increasing sophistication of modern malware. 
Most new malware families found in the wild employ some kind of functionality 
to avoid or harden detection by traditional security measures. Examples range 
from rather simplistic attempts to disable known security software upon 
infection; over polymorphism and metamorphism techniques to alter and obfuscate 
the executable binaries of malware in order to harden detection by 
signature-based approaches; up to more sophisticated behavioral obfuscation 
techniques, such as mimicry attacks, that aim at altering the runtime behavior 
to trick behavior-based detection approaches~\cite{You2010}.


One challenge of malware detection research is thus the new threat of stealthy 
and obfuscated malware; and how to counteract their attempts to avoid detection 
and remain ``below the radar''. We contribute towards this goal with a novel 
behavior-based malware detection methodology which we show to be less prone to 
circumvention by obfuscation mechanisms. The idea is to discriminate malicious 
from benign processes by analyzing their behavior \emph{in terms of 
induced quantitative data flows between system resources}. We interpret the 
execution of system calls, e.g. a process calling the Windows API 
\emph{ReadFile} function to read data from a file, as causing a quantifiable 
flow of data from one system entity, in this example a file, to another entity, 
in this example a process.

We aggregate the set of data flow events in a system within a specific time 
interval into so-called quantitative data flow graphs (QDFGs). These represent 
the interaction between all system entities within this time frame. QDFGs are 
abstractions of a system's behavior in terms of data flows, and thus can be 
used for behavior-based malware detection.
On the basis of this model our approach aims at identifying QDFG nodes that refer to potentially malicious processes. 
To do so, we use metrics 
inspired by research done in the area of social network analysis, to profile typical data flow behavior of benign and malicious processes, and then use these profiles to 
train a machine learning classifier. 

In contrast to related work on graph-based malware 
detection~\cite{Wuechner2014,Christodorescu2008,Fredrikson2010b,Fredrikson2011},
we do not rely on fixed detection patterns and expensive subgraph isomorphism 
checks. Instead, we perform approximate similarity comparison of unknown 
process behavior with a more flexible metric-based quantitative data flow 
model. By this, in contrast to isomorphism-based approaches that are challenged 
if malware does not exactly match defined patterns or models, we 
are able to detect unknown or obfuscated malware. In contrast to recently published metric-based approaches~\cite{Jang2014mal,Mao2014centrality} we incorporate quantitative data flow aspects into our model which we show to provide better detection precision and superior obfuscation resilience, as well
as novel features (which we call \emph{local}).

\textbf{Contributions}: \textbf{a)} To the best of our knowledge, we are the first to combine \emph{quantitative} data flow tracking with machine learning for checking for behavioral similarity of processes in the context of malware detection. \textbf{b)} Our experiments demonstrate the utility of \emph{quantitative} data flow aspects for detection precision. In particular we show that the consideration of quantities in data flow graphs can effectively halve false positive and false negative rates. \textbf{d)} Our evaluations indicate that our approach is more robust against common types of behavioral obfuscation than approaches that build on raw system calls such as n-gram based approaches and \textbf{e)} We show that we are able to detect samples from unknown malware families with good accuracy.

\textbf{Organization}: We recap an abstract QDFG model from the literature in \S\ref{sec:preliminaries}. We present graph metrics and their semantic relevance in terms of malware detection in \S\ref{subsec:features}; describe the training phase in \S\ref{subsec:training}; and discuss the detection procedure in \S\ref{subsec:detection}. We evaluate effectiveness, obfuscation robustness, and efficiency in \S\ref{sec:evaluation}. We put our approach in context in \S\ref{sec:related-work}. We conclude with a discussion of capabilities, limitations, and future work in \S\ref{sec:conclusion}.

\vspace{-0.5em}
\section{Preliminaries}
\label{sec:preliminaries}
\vspace{-0.2em}

We first recap some preliminaries. For the subsequent sections we assume a basic understanding of the Windows NT operating system architecture.

\subsection{Quantitative Data Flow Model}
\label{subsec:model}

We study the identification of potentially malicious processes in a system by analyzing quantitative data flow graphs (QDFGs) that represent a system's data flow activities within a certain period of time. 
To this end, we use a slightly simplified generic quantitative data flow graph model from the literature~\cite{Wuechner2014}. This model uses QDFGs  to capture all aggregated and quantified data flows between interesting entities in a system, such as processes, files, or sockets. These are represented by nodes ($\overline{N}$). Labeled directed edges ($\overline{E}$) between two nodes intuitively reflect that there has been a transfer of a certain amount of data.

A QDFG is a graph in the set $\mathcal{G} = \overline{N}\times\overline{E}\times\overline{A}\times ((\overline{N} \cup \overline{E}) \times \overline{A} \rightarrow \mathit{Value}^{\overline{A}})$, where $\overline{N}$ denotes the set of all possible nodes, $\overline{E}\subseteq\overline{N}\times\overline{N}$ the set of possible edges between two nodes, and a set of labeling functions $((\overline{N} \cup \overline{E}) \times \overline{A}) \rightarrow \mathit{Value}^{\overline{A}}$ assign defined values from the set $\mathit{Value}^{\overline{A}}$ to an attribute $a\in \overline{A}$ of a node or an edge. These labeling functions are needed to annotate nodes and edges with additional information such as amount of transferred data ($size \in \mathbb{N}$) or corresponding set of time stamps ($time \in 2^\mathbb{N}$).

QDFGs are incrementally built on the basis of data flow relevant system events, e.g. functions to read data from a file or to write data to a socket. These events are modeled as a set $\mathcal{E}$. In an actual system, they are intercepted by runtime monitors which interpret the data flow semantics and perform corresponding graph updates such as the creation or modification of nodes or edges. 
One QDFG $G = (N,E,A,\lambda)\in\mathcal{G}$ describes all data flow activities of a system that happened during a certain time interval. The labeling function $\lambda$ maps attributes of an edge or a node to their assigned values.

Events $(src,dst,size,t,\lambda)\in\mathcal{E}$ represent transfers of $\mathit{size}\in\mathbb{N}$ units of data from a node $src\in\overline{N}$ to a node $dst\in\overline{N}$ with a timestamp $t \in \mathbb{N}$ and a labeling function $\lambda$ for additional information on the corresponding data flow. To ease presentation, we will not always cleanly distinguish between an event and its corresponding edge.
We are not interested in exactly which event causes which amount of data flow between two system entities. We thus simplify our model by aggregating semantically related data flows between pairs of nodes through summation of the $\mathit{size}$ attribute of the respective edges rather than creating one distinct edge per event. 

Before formally defining the corresponding graph update function, triggered by the execution of a data flow related event, we first need to introduce some auxiliary notations and syntactic sugar:
For  $(x,a) \in  (N\cup E) \times\overline{A}$, we define $\lambda[(x,a) \leftarrow v] = \lambda'$ with $\lambda'(y) = \begin{cases} v & \mathit{if\ } y \in dom(\lambda) \\ \lambda(y) & otherwise \end{cases}$. 

We furthermore introduce some syntactic sugar for updating labeling functions: 
{\small
$\lambda[(x_1,a_1) \leftarrow v_1; \ldots ;(x_k,a_k) \leftarrow v_k]  = 
( \dots (\lambda[(x_1,a_1) \leftarrow v_1]) \dots)  [(x_k,a_k) \leftarrow v_n].
$}

Correspondingly we denote the composition of two labeling functions by:{\small
$$\lambda_1 \circ \lambda_2 = \lambda_1 [(x_1,a_1) \leftarrow v_1; \ldots ;(x_k,a_k) \leftarrow v_n]$$}
where $v_i =\lambda_2(x_i,a_i) $ and $(x_i,a_i) \in \mathsf{dom}( \lambda_2)$.
Finally, the QDFG update function $\mathit{update}:\mathcal{G}\times\mathcal{E}\rightarrow\mathcal{G}$ is formally defined in Figure~\ref{eqn:update}.

\begin{figure}[t]
	\small
	\begin{flalign*}
	&update(G,(src,dst,s,t,\lambda')) = \\
	&\begin{cases}
	\left(\begin{array}{@{}l@{}}N,\\E,\\ A \cup \mathsf{dom} (\lambda'), \\ \lambda\left[\begin{array}{@{}l@{}}(e,size) \leftarrow \lambda(e,size)+s;\\(e,\mathit{time}) \leftarrow (\lambda(e,\mathit{time}) \cup \{t\})\end{array}\right] \circ \lambda' \end{array}\right) & 
	\mathrm{if\ }
	\begin{array}{l}
 	e \in E
	\end{array}\\
& \\
	\left(\begin{array}{@{}l@{}}N \cup \{src,dst\},\\E \cup \{ e \},\\ A \cup \mathsf{dom} (\lambda'), \\ \lambda\left[\begin{array}{@{}l@{}}(e,size) \leftarrow s;\\(e,time) \leftarrow \{t\}\end{array}\right] \circ \lambda' \end{array}\right) & 	
	\mbox{otherwise}
	\end{cases}\\
	&\mathrm{where\ } e=(src,dst) \mbox{ and } G = (N,E,A,\lambda)
	\end{flalign*}
	\normalsize
	\vspace{-1em}
\caption{Graph update function \label{eqn:update}}\vspace{-1.5em}
\end{figure}

For later definitions of node features, we need to introduce auxiliary functions. Function $pre: \overline{N} \times \mathcal{G} \rightarrow 2^{\overline{N}}$ computes all immediate predecessor nodes of a node of the graph. Functions $in,out: \overline{N} \times \mathcal{G} \rightarrow 2^{\overline{E}}$ compute the set of incoming and outgoing edges of a node.

\subsection{Windows Instantiation}
\label{subsec:windows_instantiation}

To instantiate the abstract QDFG model for real-world malware detection, we need to map it to resources and events in actual execution environments. 
In this paper, the execution environment is that of typical Windows operating systems.

We identified a set of system resources that are relevant for malware data flow behavior: \emph{Processes} interact with all other relevant system entities in a way that they are either sources or sinks of flows from or to other \emph{Registry}, \emph{Socket} or \emph{Process} nodes. To type these nodes we introduce a special $\mathit{type} \in \overline{A}$ attribute: Process nodes have \emph{type} \emph{P}, File nodes \emph{F}, Socket nodes \emph{S}, URL nodes \emph{U}, and Registry nodes \emph{R}.

In addition to \emph{entities}, we also need to map all data flow relevant \emph{events}. These are all Windows API functions that lead to a flow of data between the above system entities. This includes functions to interact with resources from the file system like \emph{ReadFile} or \emph{WriteFile} to functions to send or receive data to or from a socket like the Winsock \emph{recv} and \emph{send} functions. To give an intuition how the data flow semantics of such functions is formally modeled, we present two sample function definitions:

{\small
\begin{itemize}[leftmargin=.4cm]
	\item \textbf{ReadFile} Using this function a process reads a specified amount of bytes from a file to its memory. \emph{Relevant Parameters}: Calling Process ($P_C$), Source File ($F_S$), ToReadBytes ($S_R$). \emph{Mapping}: ${(F_S,P_C,S_R,t, \lambda(F_S,{size}) :=}{\lambda(F_S,{size}) +  S_R))\in\mathcal{E}}$.
	\item \textbf{WriteFile} Using this function a process can write a specific number of bytes to a file. \emph{Relevant Parameters}: Calling Process ($P_C$), Destination File ($F_D$), ToWriteBytes ($S_W$). \emph{Mapping}: ${(P_C,F_D,S_W,t, \lambda(F_D,{size}) :=}{\lambda(F_D,{size}) + S_W))\in\mathcal{E}}$.
\end{itemize}
}

For brevity's sake we only presented two functions here to demonstrate the general procedure of mapping concrete system events to abstract events of the QDFG model. A more comprehensive list can be found for instance in~\cite{Wuechner2014}.

\begin{figure}[t]
\begin{center}
	\includegraphics[width=0.9\textwidth]{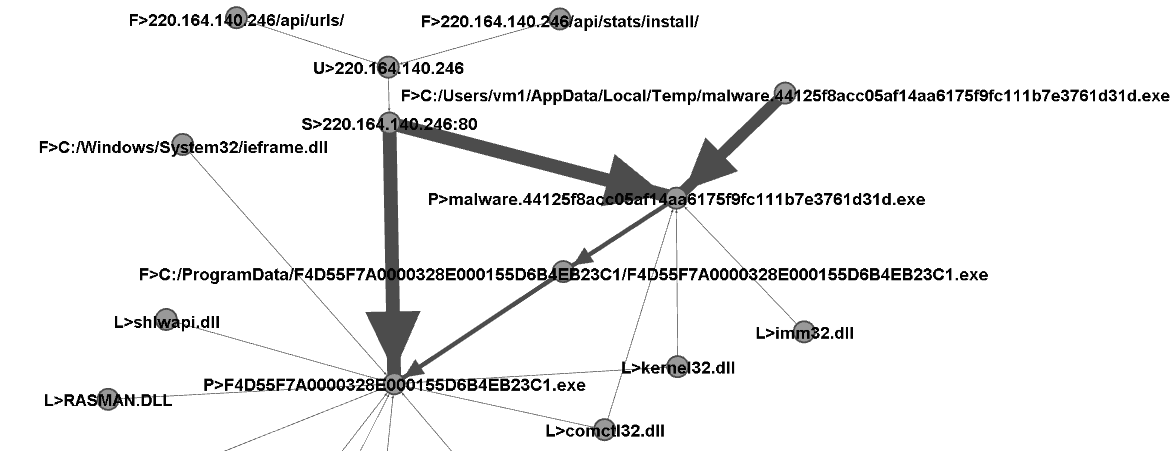}
	\caption{Excerpt of QDFG for a system infected with Cleaman.}
	\label{fig:QDFG_cleaman}
	\vspace{-2em}
\end{center}
\end{figure}

To motivate the utility of this model for malware detection, Figure~\ref{fig:QDFG_cleaman} exemplarily visualizes a typical malware QDFG, built by applying the previously mentioned model on the intercepted activities of an executed Cleaman trojan. The size of the edges in the graph visualization represents the relative amount of transferred data with respect to all other edges of that graph. The type of the nodes is denoted by the first letter of the node label. By focusing on the part of the graph with the highest amount of transferred data one can easily spot the core malign activities of the analyzed malware, i.e. self-replication, or download and execution of additional malicious payload from a remote server.

\section{Approach}
\label{sec:approach}

Our core idea is to learn statistical profiles for benign and malicious nodes in QDFGs that represent known infected and non-infected systems. We later use these profiles for matching feature sets of unknown processes against them. 

\begin{figure}[h]
	    \vspace{-1em}
	    \centering
	\includegraphics[width=0.65\textwidth]{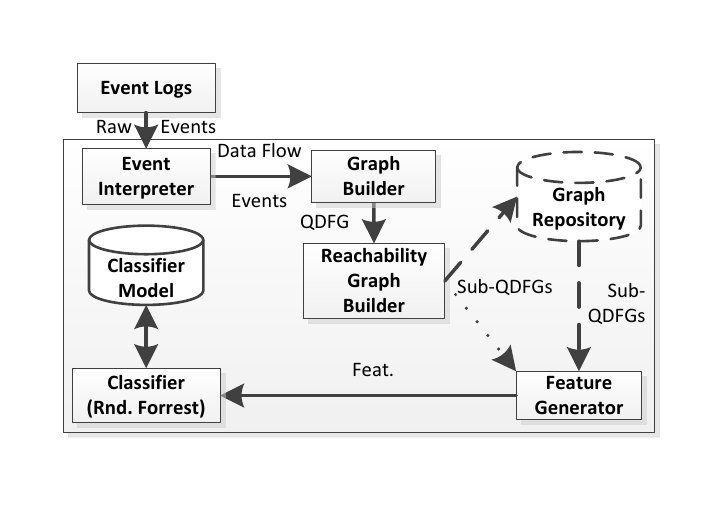}
		    \vspace{-2em}
	\caption{Architecture}
	\label{fig:architecture}
		    \vspace{-1.5em}
\end{figure}

The overall architecture is depicted in Figure~\ref{fig:architecture}. Dashed lines mark components and interactions that are only used in the training phase. Dotted lines refer to the ones only relevant for detection.

\subsection{Features}
\label{subsec:features}
Like others \cite{Jang2014mal,Mao2014centrality} we see a strong analogy between social networks and (Q)DFGs and hence use graph characteristics inspired from social network analysis~\cite{Okamoto2008,Brandes2001}. Nodes in a social network typically represent communicating entities, and the edges between them their interaction in form of exchanged messages or friendship relations. Analogously, nodes in our graphs represent system entities and the edges between them their interaction in form of data flows. 

The following features were selected using both an inductive and a deductive approach. The inductive selection was done based on a preliminary analysis of graphs of a small set of malware-infected systems where we applied several standard metrics from statistics and graph theory and tried to correlate malware activities with the applied metrics. The deductive selection was performed through an analysis of standard graph metrics. For each analyzed feature we tried to correlate its intuition with typical malware behavior or properties. For instance, malware that tries to infect other processes or files results in a high connectivity of the corresponding node to certain types of other nodes.

Features are functions $\phi \in \Phi$ that map a QDFG node to a real number:
$\phi: \overline{N} \times \mathcal{G} \to \mathbb{R}.$ 
We enrich the QDFG model to store the value of features as attributes of nodes $n$ in graph $G$, so $\lambda(n,\phi) = \phi(n,G)$.
Additionally, we distinguish between two basic types of graph features, \emph{local features} ($\Phi_l$) and \emph{global features} ($\Phi_g$). Local features have a single-hop scope. This means that they only capture the relationship of a node with its direct neighbors. Global features in contrast have a multi-hop scope and represent relationships of one node with all other nodes of a graph.

As opposed to recently published metric-based approaches~\cite{Jang2014mal,Mao2014centrality}, that also exploit graph-theoretical properties to derive discriminating features, we take into 
account quantitative data flow aspects by exploiting the additional information given by the weighted edges of QDFGs.


\subsubsection{Local Features ($\Phi_l$)}

To define the features, we need some auxiliary notation. Function $d_\psi: (\overline{N} \times \overline{N} \times \mathcal{G}) \rightarrow \mathbb{R}$ with $\psi$ returns the shortest path between two nodes in a graph, where $\psi: \overline{E} \rightarrow \mathbb{R}$ with $\psi(e) = \lambda(e,size)$ defines the edge distance, or cost, i.e. the amount of data transferred via this edge.

\begin{enumerate}[leftmargin=.4cm]
\item \textbf{Entropy} $\phi^1 \in \Phi_l$ 
computes the normalized entropy of the distribution of edge feature values such as size, event count, or sensitivity of all outgoing edges of a process node $n \in N$. The entropy captures the uniformity of the distribution of percental flows, number of contributing events, or relative sensitivity of all outgoing edges of a node $n$. 

\emph{Rationale}: Viruses like Parite infect other executable binaries or processes by injecting or appending their own binary image. The respective subgraphs tend to have a comparably uniform distribution of specific features of outgoing edges, because the majority of triggered events by that malware are targeted at the infection with roughly the same size of the events as consequence of them relating to reading or writing the same binary image.
 
\emph{Computation}: Let $\overrightarrow{s} = (s_1, \dots, s_k)$  and  $e_i \in out(n)$ in  $s_i = \frac{\psi(e_i)}{\sum\limits_{e' \in out(n)}\psi(e')}.$ 
Then we define: $ \phi^1(n,G) := \mathit{NE} ( \overrightarrow{s})$ where $\mathit{NE}(\overrightarrow{s}):=\frac{-\sum\limits_{i=1}^k s_i * \log(s_i)}{\log(k)}.$

\item \textbf{Variance} $\phi^2 \in \Phi_l$ 
expresses the statistical population variance of the distribution of a certain edge feature for all outgoing edges of a node $n \in N$. 
A low statistical variance indicates that most of the elements of the distribution elements are close to the statistical mean, whereas a high variance indicates a spread of elements. Due to its similar focus on uniformity of underlying input distributions, the variance feature is closely correlated with the entropy feature. First evaluations indicated, however, that the entropy metric performed comparably badly if a node has only a few outgoing edges, but better for larger sets of outgoing edges. The variance metric seemed to exhibit exactly the inverse characteristics.

\emph{Rationale}:
		The motivation is similar to that for entropy: malware often exhibits outgoing edge distribution characteristics different from benign ones. 
		
\emph{Computation}:
{\small
		$$
		\phi^2(n,G) :=  \frac{\sum_{e \in out(n)}{\left(\psi(e) - \frac{1}{|out(n)|} \sum_{e' \in out(n)}{\psi(e')}\right)}}{|out(n)|} 
		$$}

\item \textbf{Flow Proportion} $\phi^3_{t} \in \Phi_l$
captures the proportion of a certain type of outgoing data flows of a node $n \in N$ w.r.t. all outgoing flows of that node. The type of a flow is determined by the target node's type of the outgoing edge. We define different variants of the proportion feature that consider different edge attributes.

\emph{Rationale}:
Malware processes often exhibit different flow proportion characteristics than goodware. Examples include ransomware or virus processes that have an irregularly high percentage of outgoing edges that point to file nodes, as they either encrypt several sensitive files, or infect all executable binary files on the hard disk. 
		
\emph{Computation}: Let $t \in \{\mathit{Process, Registry,  File, Socket}\}$.
{\small		$$
		\phi^3_{t}(n,G) := \frac{\sum\limits_{e = (src,dst) \in out(n), \lambda(dst,type)=t }{\psi(e)}}{\sum\limits_{e \in out(n)}{\psi(e)}}
		$$}

\end{enumerate}

\subsubsection{Global Features ($\Phi_g$)}

Global features represent the relation between one node and---possibly all---other nodes of a graph. In contrast to local features, capture the importance of one node within the overall graph. Note that a crucial feature
of global features is the fact that the weight of edges (given by
the size of data flows between them) is considered when computing the shortest
path between nodes (given by the function $\psi(e)$).

\begin{enumerate}[leftmargin=.4cm]
\item \textbf{Closeness Centrality} $\phi^4 \in \Phi_g$
for a node $n \in N$ represents the inverse of that node's average distance to all other nodes of the same graph. A high closeness centrality indicates that the respective node is closely connected to all other graph nodes~\cite{Okamoto2008}.

 \emph{Rationale}: High connectivity with other nodes indicates a node manipulating or infecting other system resources like processes or executable binaries. Such behavior is typical for viruses like Parite that replicate by infecting other processes and binaries. This leads to a close connectivity of the corresponding malware process node with other process and binary file nodes.
 
 \emph{Computation}:
{\small		$$ \phi^4(n,G) := \frac{|N|-1}{\sum\limits_{n'  \in N \setminus{\{n\}}}{d_{\psi}(n,n',G)}} $$}

\item \textbf{Betweenness Centrality}  $\phi^5 \in \Phi_g$
of a node $n \in N$ represents the relative portion of all shortest paths between all possible pairs of nodes of a graph that pass through that specific node $n$. A high betweenness centrality means that one specific node is part of a multitude of ``communications'' between nodes~\cite{Okamoto2008}.

 \emph{Rationale}:
		This metric captures how often a process is part of a multi-step interaction or data flow between other system resources. This is useful to identify malware aiming at man-in-the-middle attacks to e.g. intercept the communication of a benign process with a socket, or to manipulate the information that a benign process reads into memory, to e.g. infect that process with malicious code at runtime. 

 \emph{Computation}: The function $sp(x,y,G)$ returns the number of shortest paths between the nodes $x$ and $y$ in a graph $G$; $sp_{z}{(x,y,G)}$ the ones that pass through node $z$.
	{\small	$$ \phi^5 (n,G) := \sum\limits_{n',n'' \in N : n \not{=} n' \not{=} n''}{\frac{sp_{n}(n',n'',G)}{sp(n',n'',G)}} $$}

\end{enumerate}

\subsection{Training and Model Building Phase}
\label{subsec:training}
We can now establish statistical profiles for the discrimination between benign and potentially malicious process nodes in a graph. 
A concrete instantiation of this training procedure with real-world data will be discussed in \S\ref{subsec:effectiveness}.
The training procedure consists of four activities: i) event log generation; ii) graph generation; iii) feature extraction; iv) classifier training. 

\subsubsection{Event Log Generation} 
Using a user mode Windows API monitor from the literature~\cite{Wuechner2012}, we log a defined amount of calls of processes to the Windows API to capture the activity and interaction of all processes within a system for a certain period of time. The monitor intercepts process calls to the Windows API and stores data flow relevant information like event name, parameter values, and name of the issuing process along with additional context information like a time-stamp to an event log.


\subsubsection{Graph Generation}
We then extract the data flow related information from the event logs. This is done by an \emph{event data flow interpreter} component that maps raw events to the semantic model discussed in \S\ref{subsec:windows_instantiation}. The \emph{graph builder} then generates one QDFG per event log as described in \S\ref{subsec:model}. In order to reduce noise, instead of storing the complete QDFG for training, we generate so-called reachability graphs for all process nodes in the base QDFGs. Such reachability graphs contain all nodes and edges that are directly or indirectly connected to the starting node. By this means we ensure, that the training graphs only contain activities that are actually triggered by a certain process or of processes that it directly or indirectly influenced, ignoring all activities that are conducted by non-related processes.

\subsubsection{Feature Extraction}
The \emph{feature extractor} computes all graph features from \S\ref{subsec:features} for all process nodes of the graphs in the training graph repository. This yields a set of feature values for different benign and malicious process. Recall that we labeled the known malicious process and thus also the corresponding graph nodes. We are hence able to label the resulting process node feature sets as belonging to a known malicious/benign process, which is a necessary precondition for later using a supervised machine learning algorithm.



\subsubsection{Classifier Training}

After the feature extraction phase, we feed the obtained features into a machine learning algorithm for training.

For this we construct a feature vector for each process node of the training set, with the elements of the vector being the considered QDFG metrics (see~\S\ref{subsec:features}), together with one label element representing the known classification (benign or malign) of the respective process. Each feature vector is thus of size $|\Phi|+1$.

Note that we only compute feature vectors for process nodes as we are solely interested in determining whether a specific process that originated from a executed binary is malicious or not; we thus do not classify a binary itself, but its runtime representation, i.e. the respective process or its children.


Considering the high number and diversity of the value space of the selected training features we need a machine learning algorithm that is robust towards training set diversity and scales well with respect to the number or training features. Initial attempts to use simple classifiers like naive Bayes yielded poor performance in terms of detection precision. We hence explored more complex algorithms like support vector machines and meta-learners. Particularly good results were achieved with the Random Forest (RF) algorithm. RF is a meta- or ensemble-learner, which means that it uses several distinct, potentially imprecise, classification models and merges their decisions to form a more precise combined decision. RF constructs many individual decision trees, called decision forest, based on random selection of limited feature subsets of the feature space. 


\subsection{Detection Phase}
\label{subsec:detection}

We now have a classifier that can predict the class (malicious or benign) of an unknown process node based on its characteristic local and global graph features. In a nutshell for the detection phase we thus only need to build a graph of a potentially infected system at runtime based on captured events, compute the characteristic features for each process node in the graph, and match the resulting feature set against the classifier.

Like for the training phase, we intercept relevant system events at runtime, interpret them in terms of their data flow semantics, and then build the corresponding (reachability) QDFGs for each process. We then compute the characteristic feature sets for the process nodes of the generated reachability graphs and match them against the classifier, using the classification model that was generated as result of the training phase. 


Consequently, all process nodes of these reachability graph are classified into benign or potentially malicious ones.

\section{Evaluation}
\label{sec:evaluation}

We implemented the detection framework and captured activities of a representative and diverse set of known benign and malicious software to assess the effectiveness and efficiency of our approach.

\subsection{Prototype}
\label{subsec:prototype}

Our prototype is a distributed system as shown in Figure~\ref{fig:architecture}. 
We used and extended a user mode Windows API runtime monitor~\cite{Wuechner2012} to intercept system activities relevant in terms of data flows of all processes running within the evaluation system. 
For the training phase we used the Random Forest implementation of the Weka machine learning framework~\cite{Hall2009}, configured to build a forest of 10 distinct decision trees using the a random feature subsets.

To be able to analyze a large body of malware samples it was necessary to automate the different analysis steps. For this purpose we customized the open-source malware sandbox framework Cuckoo~\footnote{\url{http://www.cuckoosandbox.org/}} by replacing its function call hooking module with our hooking module. 
Each Cuckoo sandbox VirtualBox instance was running a clean installation of Windows 7 SP1 and assigned two 2,4GHz cores and 2GByte of RAM. 
The generation of the QDFGs, computation of the corresponding graph features, and the classification of process samples performed on a 2,8GHz quadcore i7 system with 8GByte of RAM.

\subsection{Effectiveness}
\label{subsec:effectiveness}
As data source for our experiments we used 6994 different known malicious programs and 513 different known benign applications.

The malicious program samples were taken from a subset of the Malicia malware data set, i.e. all samples that were executable in the considered evaluation environment, that comprises of real-world malware samples from more than 500 drive-by download servers~\cite{Nappa2013}. The respective malware set consists of samples from 12 malware families, including families like zeus, spyeye, and ramnit. 

The goodware set was composed of a selection of popular applications from \url{http://download.com} and a wide range of standard windows programs, including popular email programs like ThunderBird, browsers like FireFox, video and graphics tools like Gimp, or VLC Player, and security software like Avast. 

We generated about 7500 event logs and converted them into QDFGs, each capturing activities of the sandbox machines for a time interval of 5 minutes. 


With this data and the procedure explained in the previous section we obtained a total of 8648 (i.e. 1654 goodware and 6994 malware) QDFG feature sets. The reason for the set of goodware features being bigger than the set of executed goodware samples is that for each execution of a goodware sample we did not only capture the behavior of the goodware sample itself, but also the interaction with all simultaneously running standard Windows processes. 

To evaluate the detection performance of our approach on the obtained feature set we first performed ten times a 10-fold cross validation test. For this tests we split the entire feature set into two parts, using 90\% of the set for training and the remaining 10\% for testing. The sets were randomly generated and the splitting repeated 10 times for each test to limit bias from specific set compositions. For each run we built a classification model on basis of the training data and used it for classifying the remaining test set. 
To avoid training bias due to unbalanced feature sets we in addition applied a SMOTE oversampling~\cite{Chawla2002} on the training sets to approximately balance the distribution of malware and goodware samples.
To express the effectiveness, we computed the following quality metrics. True positives (TP) refer to malware samples (MW) that have been correctly classified as malicious, true negatives (TN) to goodware samples (GW) that were correctly classified as benign, false positives (FP) to goodware samples incorrectly classified as malicious, and false negatives (FN) malware samples that were mistakenly labeled as benign:

\begin{center}
Detection Rate (DR)~:~$\frac{TP}{MW}$ \hspace{1em} False Positive Rate (FPR)~:~$\frac{FP}{GW}$ \\
 Precision~:~$\frac{TP}{TP + FP}$  \hspace{1em} F-Measure~:~$\frac{2 * TP}{2 * TP + FP + FN}$
\end{center}

\begin{table}[t]
		{\small
	\centering
		\begin{tabular}{|p{2.2cm}||p{1.7cm}|p{1.7cm}|p{1.7cm}|}
			\hline
			& a) Real & b) Fixed & c) Random\\
			\hline
			\hline
			\doublecell{Avg. Det. Rate\\(Std. Dev.)} & \doublecell{98.01\%\\($\sigma=0.51\%$)} & \doublecell{98.00\%\\($\sigma=0.57\%$)} & \doublecell{95.23\%\\($\sigma=0.90\%$)}\\
			\hline 
			\doublecell{Avg. FP Rate\\(Std. Dev.)} & \doublecell{0.48\%\\($\sigma=0.34\%$)} & \doublecell{0.85\%\\($\sigma=0.35\%$)} & \doublecell{1.08\%\\($\sigma=0.41\%$)}\\
			\hline
			Precision & 99.62\% & 99.32\% & 99.12\%\\
			\hline 
			F-Measure & 98.81\% & 98.65\% & 97.13\%\\
			\hline 
		\end{tabular}\vspace{1em}
	\caption{Effectiveness Quality Metrics}
	\label{tab:EffectivenessQualityMetrics}
	} \vspace{-1.5em}
\end{table}	

Table~\ref{tab:EffectivenessQualityMetrics} (a) depicts the average \emph{effectiveness quality metrics} of the cross validation experiments. As we can see, our approach at average can correctly detect 98~\% of the provided malware set with a low false positive rate of only about 0.5\%. The low standard deviations furthermore indicates a good stability of the results.

\subsubsection{Impact of Quantities} 
\label{subsec:impactQ}
To evaluate our hypothesis, that the consideration of quantities has a significant impact on the effectiveness of the classification, we performed two more tests. For the first test we replaced the real quantities associated to the edges of the QDFGs with a globally fixed value of 1. For the second test we performed the edge quantity replacement by associating varying random quantities to the edges. With this we effectively destroyed the inherent quantitative information of the QDFGs. For both experiments we again performed 10-fold cross validation tests to ensure stability of the results. Table~\ref{tab:EffectivenessQualityMetrics} (b) and (c) depicts the average detection and false positive rates for both settings.

To calculate the relative impact of quantities on the detection effectiveness we divided the false positive and false negative rate (which is the dual of the detection rate) for the fixed and randomized quantities experiment by the respective rates of the experiment with the real quantities.

As we can see, fixing the quantities to a constant value increases the false positives by a factor of 1.8 ($\frac{.0085}{.0048}$). For the randomized quantities experiment we could observe an even bigger loss of effectiveness. Here the false positives increased by a factor of 2.3 ($\frac{.0108}{.0048}$), while also the false negatives increased by a factor of 2.4 ($\frac{1-.9523}{1-.9801}$) with respect to the experiments with the actual quantities.

These observations thus support the hypothesis about the utility of quantitative information for malware detection.
To verify the statistical significance of these finding we performed a two-tailed t-test on the detection and false positive rates of the different experiments. The resulting p-values were all far below 0.01\%, which indicates a high statistical significance of our observation.

\subsubsection{Ability to Detect New Malware}
For evaluating our second hypothesis, i.e. that we are able to detect new malware, we performed an additional classification experiment. For each experiment run we split our data set into two parts. The first part, which we used for training, contained all goodware samples and the samples of all malware families except for one. The second set correspondingly contained all samples from the remaining family and was used as test set. 

With this strategy we ensured that the training set did not contain any samples from the same family that was used for testing. In consequence the classifier could not gain any knowledge about the to be classified malware family.

With this test procedure we simulated the real-world scenario that our approach faced a sample from a new malware type that was never seen before and thus could not be used for training the detection model.

\begin{figure}[t]
\begin{center}
	\vspace{-1em}
		\includegraphics[width=0.9\textwidth]{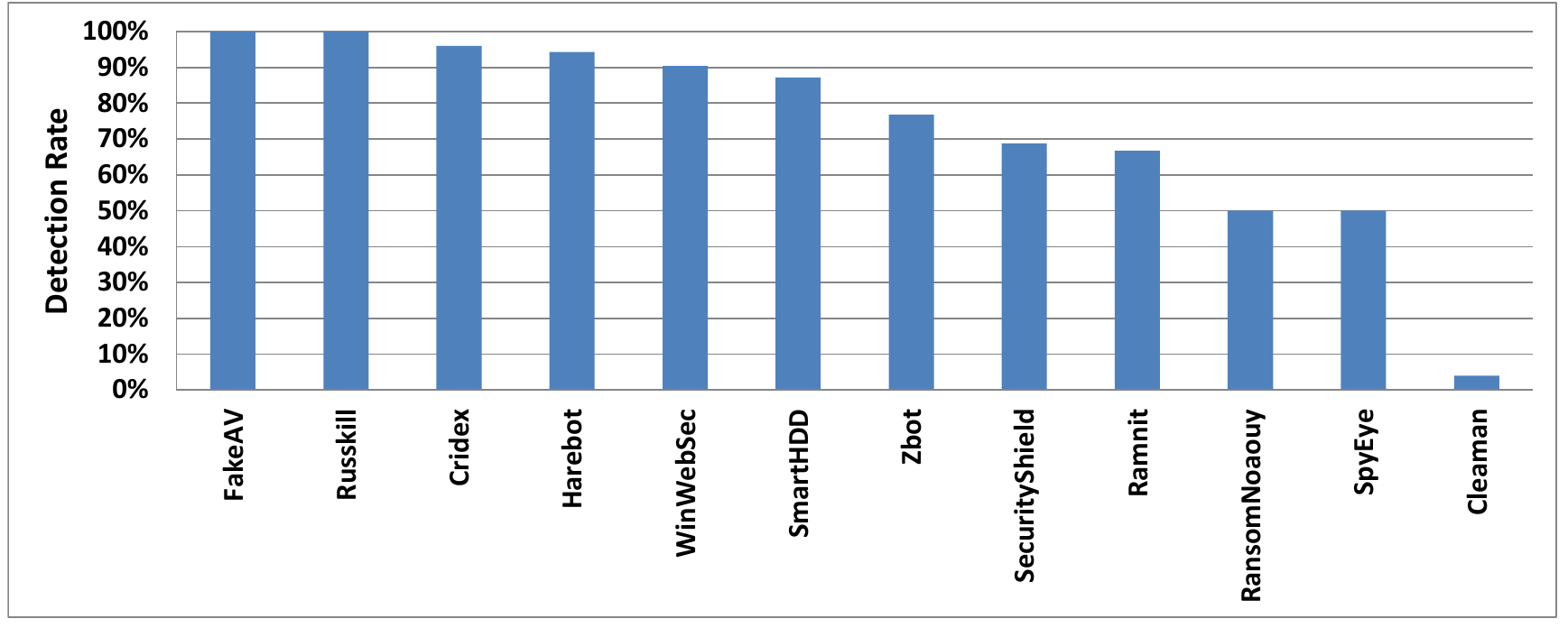}\vspace{-0.5em}
	\caption{Detectability of new malware} \vspace{-2em}
	\label{fig:evaluation_new_malware}
\end{center}
\end{figure}

Figure~\ref{fig:evaluation_new_malware} depicts the detection rates that could be achieved for the different malware families. Each bar shows the percentage of all malware samples of a specific family that could be detected using a classifier that was trained on the samples from the remaining malware families.

As we can see, our approach in all cases was able to detect samples from unknown malware families. On average our approach was able to correctly identify 73.68\% of the new malware samples; some malware families could even be classified with 100\% correctness. These results for this experiment supports our hypothesis that our approach is capable of detecting new unknown malware and goodware. Although further
investigation is needed to understand why, we speculate that this is the case because malware behavior is often not too different, even between distinct malware families.

\subsubsection{Obfuscation Resilience}

 
Approaches that obfuscate the binary image of malware through build-time code encryption and run-time decryption or code diversification barely have any influence on the detectability through behavior-based detection approaches as such code transformations typically do not alter the externally observable program behavior. In consequence, our approach is likely to be widely robust to such used code obfuscation. This assumption is supported by the ability of our approach to detect variants of malware families that were obfuscated through code transformations. Our evaluations for instance show that we were able to detect 96 of 101 variants of the Harebot trojan from our data set, which is known to employ different forms of code obfuscation.


On the other hand, if malware e.g. non-deterministically executes bogus non-malicious activities 
or randomly alters between semantically equivalent system calls to achieve the 
same behavior, it can effectively trick common behavioral approaches that base on n-grams 
profiling and re-identification as consequence on the unpredictable diverse 
resulting n-grams as we will show in the following. The same holds for most 
call-graph based approaches as 
call-graphs can be easily obfuscated by altering or reordering system calls.

Our approach is by construction more robust against call 
reordering or substitution approaches. This is because reordering of system 
calls does not alter the corresponding QDFGs, and because 
semantic substitutions of system calls typically exhibit similar data flow 
properties that result in similar QDFG updates. Moreover, the 
injection of bogus calls can change QDFGs, in particular if new edges are 
created in consequence of e.g. previously untouched system entities being read 
or written to, or if certain operations are repeated such that the edge weights are altered. 

To evaluate the absolute effects of different types of behavioral obfuscation 
techniques on the effectiveness of our approach we thus set up a series of 
additional experiments.

First, we picked a set of 100 malware and 100 goodware samples as baseline for our experiments. 
To reason about the obfuscation resilience of our approach and related behavioral detection approaches like n-gram based ones we then step-wise applied behavioral obfuscation transformations on the call traces of these samples to achieve two typical types of behavior obfuscation, namely re-ordering of calls and injection of bogus calls. We did so by applying a behavior obfuscation tool~\cite{ObfPaper} to the baseline malware set. This tools obfuscates commodity malware by randomly reordering its issued, or injecting new, system calls.

To investigate the effect of increasing degree of behavior obfuscation, we repeated these obfuscation steps 360 times with different configurations for call reordering and injection probabilities, as well as different upper bounds for to be reordered or injected calls. We represented the degree of obfuscation as Levenshtein Distance between the obfuscated and the non-obfuscated baseline call traces. More specifically, we computed the average number of insertions, deletions, and substitutions needed to transform the non-obfuscated call traces into the respectively obfuscated ones. 
To also get a relative comparison with other behavioral, raw system call based detection approaches, we conducted these experiments with measuring both, the detection effectiveness of our approach and the effectiveness of a typical behavioral detection approach based on n-grams of unordered system calls without arguments with varying n-gram sizes as e.g. discussed in~\cite{canali2012quantitative}.

\begin{figure}[t]
				\begin{center}
        \begin{subfigure}{0.8\textwidth}
                \includegraphics[width=\textwidth]{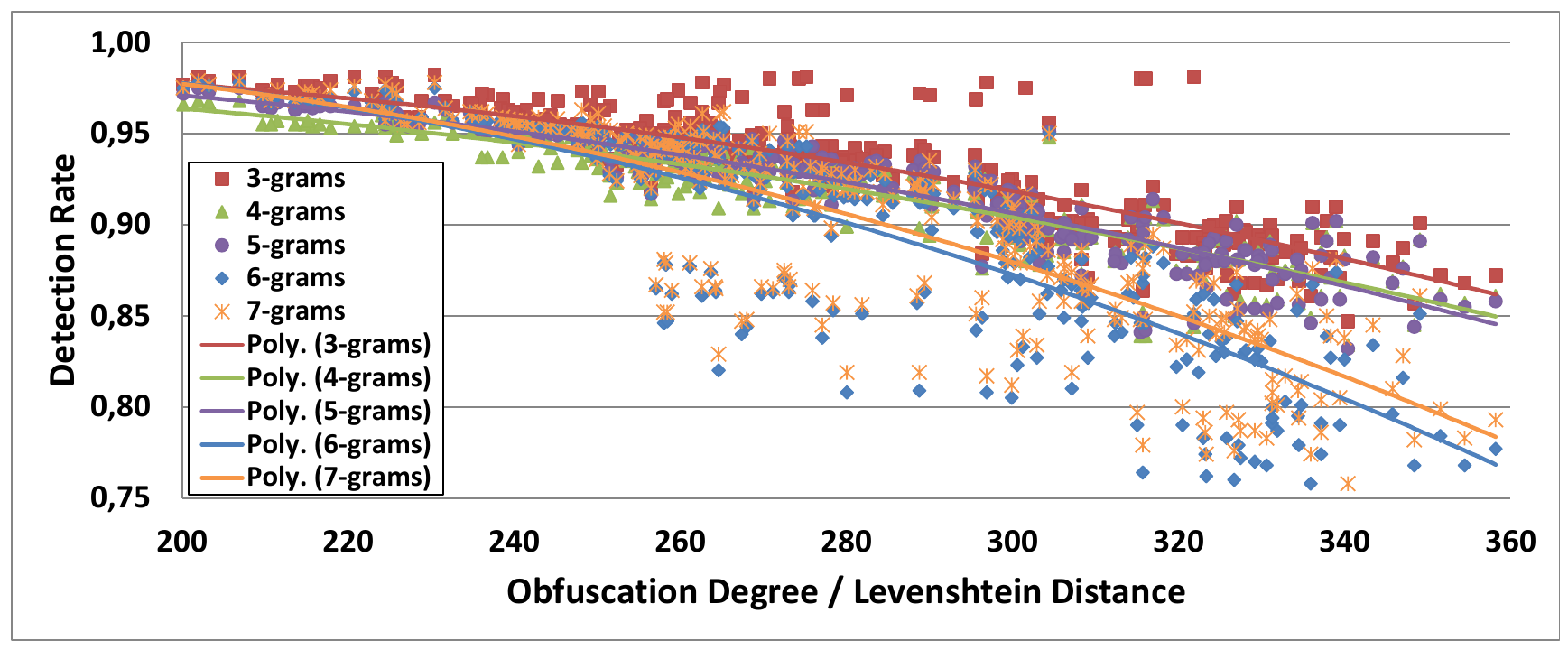}
                \caption{n-gram classifier}
                \label{fig:obfuscation_ngram}
        \end{subfigure}
        \begin{subfigure}{0.8\textwidth}
                \includegraphics[width=\textwidth]{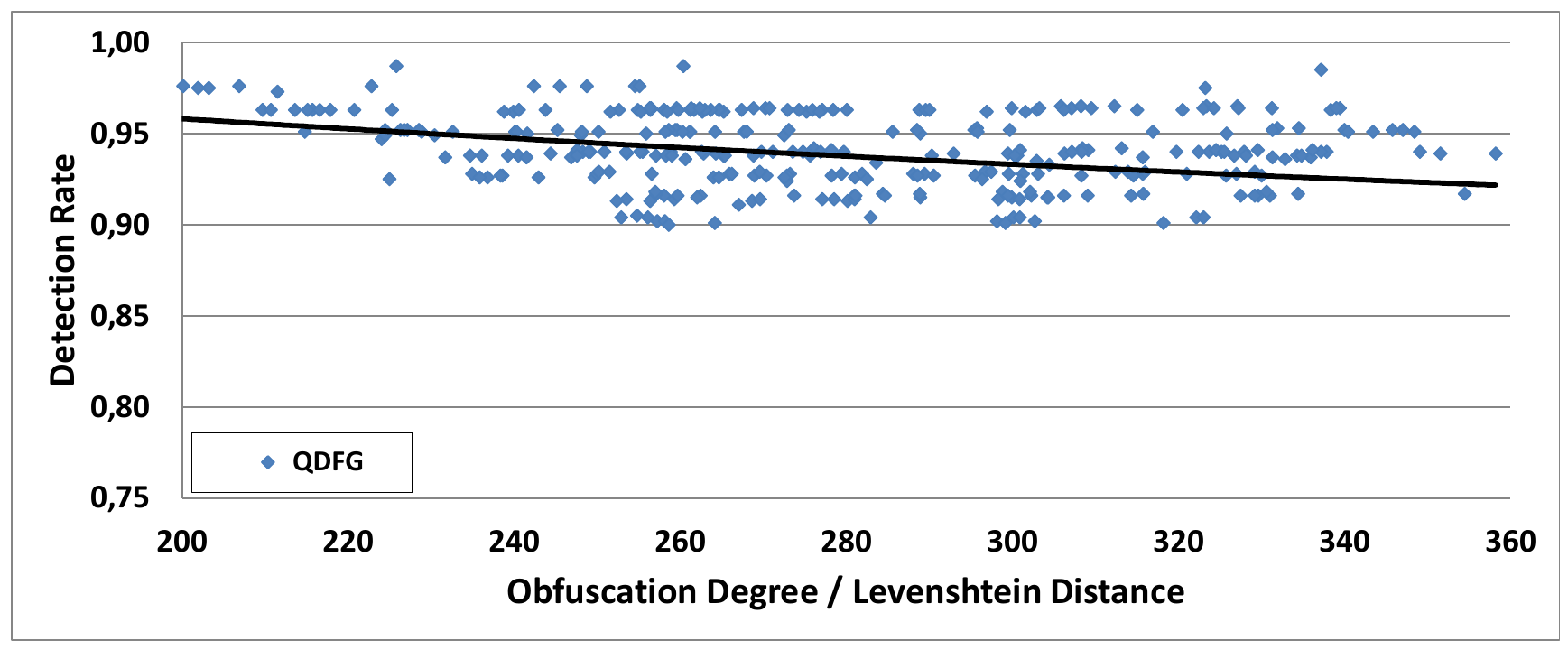}
                \caption{QDFG metric classifier}
                \label{fig:obfuscation_qdfg}
        \end{subfigure}
        \caption{Obfuscation experiments}
		\vspace{-2em}
				\label{fig:obfuscation}
				\end{center}
\end{figure}

As we can see in Figure~\ref{fig:obfuscation}, the ability of the n-gram based approach to correctly identify malicious processes significantly dropped with increasing obfuscation degree. In particular, higher-order n-grams seem to be more sensitive to behavioral obfuscation than lower-order ones. In contrast to n-gram based approaches, our approach seems to be considerably more robust and stable towards behavioral obfuscation. While the effectiveness of the n-gram approaches quickly dropped quadratically and lost prediction stability when obfuscation transformations where applied, the effectiveness of our approach at the same time remained rather stable and only slowly dropped linearly. 

In sum, our evaluation indicates that we are rather robust with respect to realistic behavior obfuscation such as random bogus call injection or reordering, whereas we could show that common n-gram based approaches are considerably challenged by such obfuscation techniques.

\subsection{Efficiency}
\label{subsec:efficiency}

The generation of Random Forest classification models took between $55.21$ and $75.38$ seconds. The size of the generated models was between $16$ and $19$ MBytes. As the training and model generation phase is only conducted once, this overhead does not contribute to the overhead during the detection phases.

\begin{figure}[!t]
\begin{center}
		\includegraphics[width=0.8\textwidth]{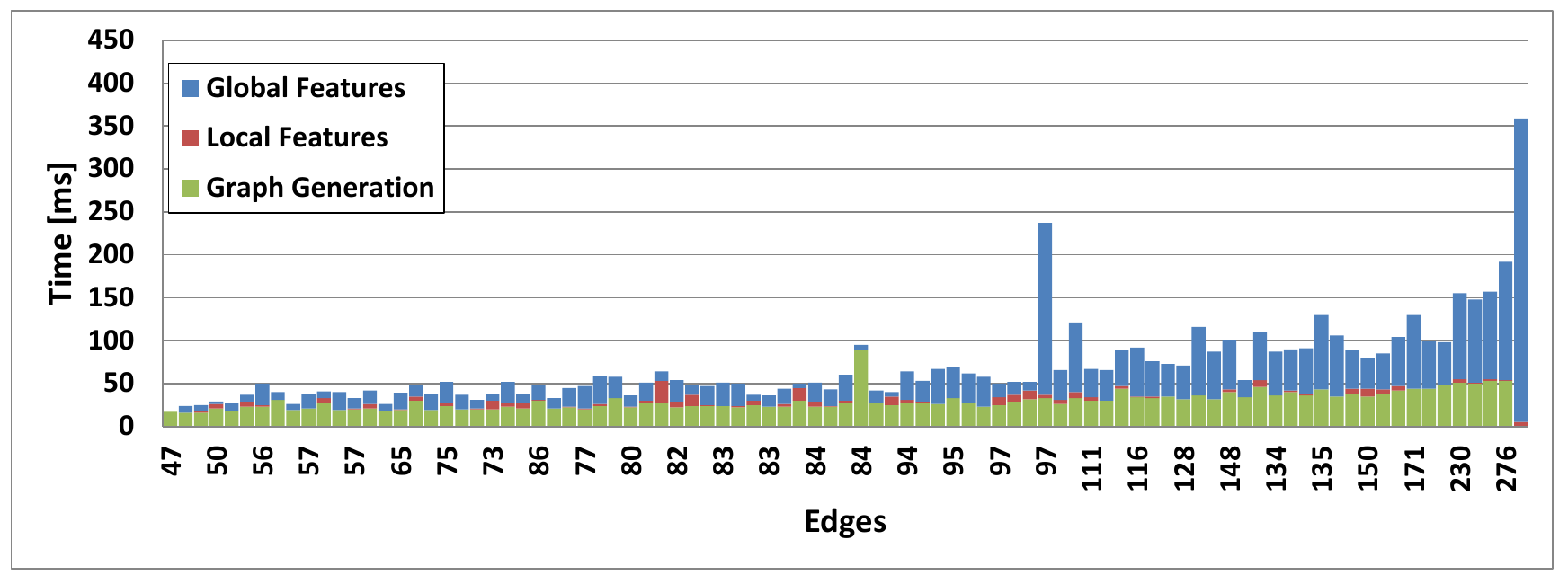}\vspace{-1em}
	\caption{Computation Time vs. Graph Complexity}\vspace{-2em}
	\label{fig:evaluation_efficiency_component}
\end{center}
\end{figure}

As we can see in Figure~\ref{fig:evaluation_efficiency_component} the overall detection time seems to increase quadratically with respect to the graph size. The bottom-most part of the area stack refers to the time it took to generate a QDFG from a given event log, the area on top of that indicates the time needed to compute the local graph metrics, and the top-most area expresses the time spent for computing the global metrics. This is not surprising, as most graph algorithms such the used centrality metrics have a theoretical complexity of $\mathcal{O}(n^2)$ to $\mathcal{O}(n^3)$, with $n$ being the number of nodes in the graph~\cite{Brandes2001}. On the other hand, the overhead for graph generation and computation of local features only grows linearly with respect to the graph size.
The overhead induced by matching the generated graph features against the classification model was below the evaluation precision threshold of $1$ ms and thus ignored for this analysis as it has no noticeable impact on the overall overhead.
%
The size of the QDFGs used for our evaluation ranged from $47$ to $330$ edges resulting in an overall detection time between $24$ms and $412$ms. On average, generating a QDFG, computing its local and global features, and matching it against the trained classifier took $71.213$ms.


\subsection{Summary and threats to validity}
\label{subsec:threats_to_validity}

In sum, we have shown our approach to be highly effective; to yield these results, among other things, on the basis of quantities; to be able to detect malware families for which it has not been trained; to be robust w.r.t. obfuscation -- in particular more robust than n-grams; and to be efficient.

Naturally, these results need to be put in the context of our study. The evaluation results were obtained on the basis of a proof-of-concept prototype with event logs generated in a controlled lab environment. The obtained insights might not generalize to the application of our approach in real-world settings. 

We tried to limit the risk of over-fitting the model to specific sub-sets of the training data by pro-actively diversifying the training set by selecting a wide and diverse range of popular malware and goodware samples for our training set. Furthermore, the fact that our cross-validation results show a very low standard deviation supports the assumption that we did not over-fit our models.

We executed and sampled the evaluation malware in a virtualized sandbox environment. It is known that modern malware often includes functionality to detect virtualization. 
There hence is a risk that we train our classification models on unrealistic malware behavior and thus are not able to detect their behavior in real-world scenarios.
Even though we did not explicitly address this threat, for the future we plan to generate malware event data on realistic bare-metal environments and re-run the experiments to find out if this has any effect on the performance of our approach.

Because current malware still mainly focuses on avoiding detection by signature-based approaches, it rarely employs advanced behavior obfuscation techniques. 
Although we applied as much randomization as possible for our obfuscation it can of course not be excluded that the simulations do not adequately reflect real-world obfuscation techniques, or that adversaries might come up with more complex behavioral obfuscation operators.


\section{Related Work}
\label{sec:related-work}

While behavior-based detection algorithms can still not be considered standard for commercial products, they have a long history in academia, especially in the intrusion detection domain. We do not perform a full literature survey on this topic here but focus on dynamic behavior-based malware detection.


A seminal idea of Forrest et al.~\cite{Forrest1996} for behavior-based analysis was to profile benign and malign processes on the basis of characteristic system call sequences (n-grams). This approach was later refined and combined machine learning methods to improve classification effectiveness~\cite{Lee1997,Ghosh1999,Lanzi2010,Milea2012,Wressnegger2013}. Similar ideas were also used to classify malware w.r.t. behavioral similarities~\cite{Bailey2007,Rieck2008,Rieck2011}.

Our system also intercepts and processes system calls. However, while the previously mentioned approaches directly use sequences of raw system calls, we use \emph{a data flow abstraction} of these calls. This distinction is important as approaches that base on raw system call data are often significantly challenged by malware that use behavioral obfuscation techniques~\cite{Borello2008,Sharif2008,You2010}. 
Such obfuscations for example target at breaking n-gram based approaches by reordering or randomizing system call sequences. As we base our analysis on a (quantitative) data flow abstractions of the system calls, which is independent of sequence order, our approach is, by construction, more robust to such obfuscation attempts.

Besides those detection approaches that use non-in\-ter\-pre\-ted sequences of raw system calls, a separate line of research performs intermediate interpretation steps. A common approach is to extract semantic dependencies between different system calls of a process to form characteristic profiles for known goodware and malware. 
Popular examples represent system call dependency profiles in form of data or control flow graphs~\cite{Kolbitsch2009,Park2013,Fredrikson2010b,Lee2010}, or assign high-level semantics to known graphs~\cite{Christodorescu2005,Christodorescu2008,Preda2007}. (Sub-)graphs that pertain to known malicious behavior are then used to re-identify malicious behavior of system processes at runtime. Due to the used intermediate abstraction steps such approaches also appear more robust to behavioral obfuscation attempts. Unfortunately, they are also challenged in terms of identifying previously unknown malicious behavior for which detection profile graphs have not yet been extracted.

In particular, from the data flow perspective, close to ours is the work of Park et al. \cite{Park2013}. They
construct Data Flow Graphs based on system calls, where entities are processes, 
child processes and files. Based on the DFGs of variants of a given malware 
family, they compute a common sub-graph called a \emph{HotPath}. This process 
can be repeated for sets of variants of different malware families, and the 
resulting sub-graphs can be used to classify the DFG of a given process: it 
will be a variant of a known malware family if it contains a similar sub-graph 
or goodware otherwise. Different from our work, they do not consider quantities 
in their graph representation (which we have shown to be an important 
discriminating factor) and by construction their approach is tailored to 
recognize mutations of known malware families. They discuss the robustness
of their approach against similar obfuscation techniques as the ones we
consider, but opposed to our work their approach is challenged by the injection
of arbitrary bogus system calls. 

Other approaches~\cite{Yin2007,Kirda2006} share similar drawbacks, as they depend on explicit definitions of malicious behavior. Our approach in contrast does not rely on fixed detection patterns. Due to the generic high-level nature of the used data flow graph features, it is more likely to also be able to detect new attacks and malicious behavior that deviate from the ones that were used for training. The use of statistical graph-based metrics for detection instead of fixed data flow patterns also differs from previous work on malware detection through quantitative data flow graphs~\cite{Wuechner2014}.
The main difference to related work that leverage taint analysis for anomaly-based malware detection~\cite{Bhatkar2006,Cavallaro2011} is that we leverage quantitative data flow aspects without using comparably expensive taint tracking.

The recently published work of Jang et al.~\cite{Jang2014mal} relates to our work in that they also leverage graph metrics to discriminate malware from goodware. But, in contrast to our work, they base the computation of those metrics on system call dependency graphs, while our model is based on quantitative data flow graphs. As we could show, this abstraction increases the robustness towards behavioral obfuscation which gives us better resilience than approaches that directly base on raw system calls. Mao et al.~\cite{Mao2014centrality} also leverage graph metrics on system entity dependency graphs for malware detection. Similarly, in contrast to us they do not incorporate any quantitative flow information for which we could show that it has an considerable impact on detection precision.

In sum, the main technical difference between our work and related contributions is that we leverage QDFG features rather than raw system calls or system entity dependency graphs. This makes our approach fast and robust against common types of behavioral obfuscations, and, due to the additional quantitative dimension, we achieve a good detection precision.

\section{Discussion and Conclusion}
\label{sec:conclusion}

We have presented a novel approach to perform graph metric based malware 
detection on the basis of quantitative data flow analysis. 
We intercept system calls issued by system processes, interpret them in terms 
of their data flow semantic and build quantitative data flow graphs. These are 
used to identify graph nodes that represent potentially malicious system 
processes.

To this extent, we compute sets of characteristic graph features, such as 
centrality metrics, for each process node in the graph to discriminate between 
benign and potentially malign nodes through a machine learning classifier. 
Using this classifier, trained on feature sets of known goodware and malware, 
we are able to discriminate unknown process samples.

It is difficult to objectively compare the effectiveness of different malware 
detection approaches presented in literature, due to varying evaluation 
baselines and used assessment procedures.
However with respect to closely related dynamic malware detection 
approaches~\cite{Yin2007,Kolbitsch2009,Fredrikson2010b}, our approach has 
similar or better detection effectiveness, while achieving a significantly 
better efficiency. 

In contrast to previous QDFG-based work~\cite{Wuechner2014} and related 
rule-based approaches~\cite{Yin2007}, our approach is able to detect novel 
malware samples that exhibit unknown behavior with better detection 
effectiveness and efficiency. In comparison to related metric-based 
approaches~\cite{Jang2014mal,Mao2014centrality} we could show that the 
quantitative aspect significantly improves detection precision.

Moreover, we have shown that our approach is robust to certain classes of
behavioral obfuscation: by construction the \emph{order} of system calls
is irrelevant, since they produce the same QDFGs, and more interestingly, 
random injection of system calls that potentially modify both the structure and 
the original quantities does not significantly alter the detection 
effectiveness either.

In conclusion, we showed the usefulness of quantitative data flows for malware 
detection and established a foundation for further research in the area of QDFG 
based malware detection models. 
We plan to perform further tests on the robustness of our approach, try to 
generalize our approach to non-sandbox settings, and to improve effectiveness 
through additional graph features.

\bibliographystyle{abbrv}
\bibliography{biblio_compressed} 

\end{document}